\documentclass[10pt,prl,aps,twocolumn,showpacs]{revtex4}
\usepackage{graphicx}

\begin{document}

\title{Quantum-criticality and percolation in dimer-diluted 2D antiferromagnets}

\author{Anders W. Sandvik}
\affiliation{Department of Physics, Boston University, 
590 Commonwealth Avenue, Boston, Massachusetts 02215}

\begin{abstract}
The $S=1/2$ Heisenberg model is considered on bilayer and single-layer square lattices with 
couplings $J_1$, $J_2$, and with each spin belonging to one $J_2$-coupled dimer. A transition 
from a N\'eel to disordered ground state occurs at a critical value of $g=J_2/J_1$. The systems 
are here studied at their dimer-dilution percolation points $p^*$. The multi-critical point $(g^*,p^*)$ 
previously found for the bilayer is not reproduced for the single layer. Instead, there is line 
of critical points $(g < g^*,p^*)$ with continuously varying exponents. The uniform magnetic 
susceptibility diverges as $T^{-\alpha}$ with $\alpha \in [1/2,1]$. This unusual behavior is 
attributed to an effective free-moment density $\sim T^{1-\alpha}$. The susceptibility of the
bilayer is not divergent but exhibits remarkably robust quantum-critical scaling.
\end{abstract}

\date{\today}


\maketitle

A challenging and important aspect of quantum phase transitions \cite{sachdev} is the influence 
of disorder (randomness) on the critical behavior, in the ground state as well as in the 
finite-temperature quantum-critical scaling regime \cite{chn,chubukov}. For one dimensional quantum 
spin systems a real-space renormalization group (RG) scheme \cite{ma} has rigorously established 
\cite{fisher} a strong-disorder {\it random singlet} fixed point which controls the low-energy 
behavior for any strength of the disorder. This is an unusual type of quantum criticality with dynamic 
exponent $z=\infty$, which leads to, e.g., a susceptibility diverging as $T^{-1}\ln^{-2}(T)$. The RG 
procedure has also been carried out numerically for various two-dimensional (2D)  models, although 
in this case there is no rigorous proof of its validity. In the random transverse Ising model a random 
singlet phase was found \cite{pich}, but only conventional critical points (finite $z$) were
found for SU(2) symmetric (Heisenberg antiferromagnet) systems \cite{lin}. However, the disorder leads 
to unusual properties in the ordered and disordered phases \cite{lin,lafl}, e.g., quantum 
Griffiths effects.

\begin{figure}
\includegraphics[width=6.5cm]{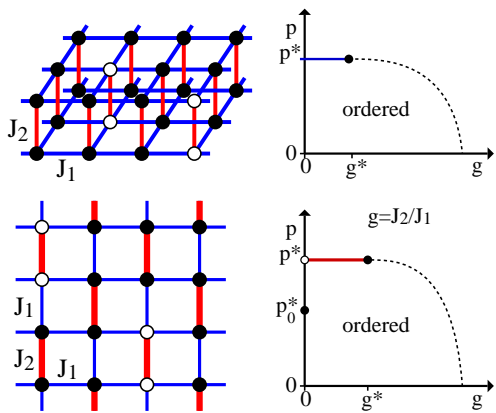}
\caption{(Color online) The lattices studied (left)
and their $T=0$ phase diagrams (right). The couplings are $J_1{\bf S}_i\cdot {\bf S}_j$ 
and $J_2{\bf S}_i\cdot {\bf S}_j$, with $J_1,J_2 >0$ (in the bilayer all inter-layer couplings are
$J_2$ and in the single layer the vertical bonds alternate $J_1$,$J_2$). Dimer dilution corresponds to 
removing $J_2$-coupled pairs---such removed dimers are indicated by open circles. For the bilayer, 
the percolating cluster is ordered on the line $(0 \le g < g^*,p^*)$, whereas the single-layer is 
quantum-critical for $(0 < g < g^*,p^*)$.}
\label{fig1}
\end{figure}

In the 2D $S=1/2$ Heisenberg model an order-disorder transition occurs at a critical 
strength of a coupling pattern favoring singlet formation on dimers, plaquettes, 
etc.\cite{chn,chubukov}. There is ample evidence from quantum Monte Carlo (QMC) studies 
\cite{troyer,matsumoto,bilayqmc} that this transition is in the 3D $O(3)$ universality class, as 
predicted theoretically \cite{haldane,chn}. According to the Harris criterion \cite{harris}, 
disorder should be relevant at this transition. The applicability of efficient quantum Monte 
Carlo methods \cite{sse}, in combination with a well developed RG scheme, makes dimerized Heisenberg 
models very well suited for exploring disorder effects on quantum phase transitions. Recent 
studies have demonstrated a variety of scaling behaviors\cite{lin,sandvik1,vajk2,sknepnek,yu,monthus}. 
The case of dilution disorder is particularly interesting, as it includes the special case 
of quantum-criticality at the classical percolation point, where geometrical and spin fluctuations 
are simultaneously divergent \cite{sandvik1,vajk2}. Here two cases of such multi-criticality 
are studied using QMC simulations---a bilayer and a dimerized single-layer model with random dilution
of dimers, illustrated in Fig.~\ref{fig1} along with schematic phase diagrams. The dramatically
different behaviors at the percolation threshold is the main focus of this Letter. The bilayer has 
a single point of quantum-criticality where the susceptibility scales to zero as $T\to 0$ 
\cite{sandvik1,vajk2}. In contrast, the single layer exhibits a line of quantum critical points 
where the susceptibility is divergent.

The 2D Heisenberg model with random site dilution has been studied extensively because of its 
relevance to Zn doped cuprate antiferromagnets \cite{carretta,vajk}. It was for a long time believed  
that site or bond dilution leads to a quantum phase transition before the geometric percolation 
point is reached \cite{wan,yasuda,chen}. However, recent QMC studies have shown that the transition 
coincides with the percolation point \cite{kato,perc} and that the percolating cluster 
is ordered \cite{perc}. The critical exponents pertaining to equal-time correlations 
($\nu$, $\beta$, and $\eta$) are therefore those of classical percolation. Other exponents 
($\alpha,\gamma, \delta$) are given by combinations of percolation exponents and the dynamic 
exponent of the spin clusters \cite{vojta}. 

The long-range order of the Heisenberg antiferromagnet on percolating clusters, which have fractal
dimensionality $d=91/48$ \cite{stauffer}, implies that other couplings have to be introduced 
in order to realize a dilution-driven quantum phase transition \cite{perc}. This has been 
explored recently with the dimer-diluted bilayer model \cite{sandvik1,vajk2}. In the pure 
Heisenberg bilayer, with intra- and inter-layer couplings $J_1$ and $J_2$, there is an 
order-disorder transition at $g=J_2/J_1 \approx 2.52$ \cite{bilayqmc} (due to the tendency 
to singlet formation across the planes). If this system is diluted by randomly removing single 
spins, order is induced in the disordered phase because moments localize in the neighborhood 
of the vacancies. These moments interact and order at $T=0$ \cite{singledil}. By instead diluting 
the two layers symmetrically, i.e.,removing dimers consisting of nearest-neighbor spins on opposite 
planes, no localized moments form and a phase transition takes place at a coupling $g$ which 
depends on the dilution fraction $p$, as shown in Fig.~\ref{fig1}. At the geometrical percolation 
point, which clearly is the same as for a single site-diluted layer, $p^* \approx 0.41$ 
(hole concentration) \cite{stauffer}, there is a multi-critical point $(g^*,p^*)$ at which the 
long-range order on the percolating cluster vanishes and the spins are quantum-critical. A critical 
coupling $g^* \approx 0.15$ and dynamic exponent $z^* \approx 1.3$ were found in two independent 
QMC calculations \cite{sandvik1,vajk2}. The generic transition for $p < p^*$ has also been studied 
in detail by Monte Carlo simulations of an analogous 3D classical Heisenberg model with 
columnar defects \cite{sknepnek}. Also here an exponent $z\approx 1.3$ was found (but 
$z^* \not = z$ is expected because of the different cluster dimensionality). 
Long-range order in the presence of quantum fluctuations on the line $(g < g^*,p^*)$, which
was believed not to be possible for a continuous order parameter \cite{senthil}, was recently 
related to the fracton dimensionality of the percolating cluster \cite{brayali}.

The question now arises as to the generality of the behavior found in the bilayer model. 
On its percolating cluster each spin has a neighbor in the opposite layer with which it correlates 
at low temperature. Magnetization fluctuations are thus quenched as $T \to 0$. 
Here the dimerized single-layer model is used to investigate the role of the bilayer symmetry 
upon dilution. Without disorder, the system has a quantum-phase 
transition, in the same universality class as the bilayer, at $J_2/J_1 \approx 2.5$. With dimer 
dilution again corresponding to random removal of $J_2$ dimers, the percolation point $p^*=1/2$ 
because the dimers are connected as a triangular lattice \cite{stauffer}. 
The $T=0$ phase diagram, outlined in Fig.~\ref{fig1}, is similar to that of 
the bilayer in that there is a finite segment of the phase boundary at $p=p^*$, terminating 
at a point $(g^*,p^*)$ beyond which the transition occurs for $p < p^*$. 
However, as will be discussed in detail below, there is a striking difference: Whereas 
the percolating cluster of the bilayer has N\'eel order for $0 \le g < g^*$ \cite{sandvik1,vajk2}, 
the percolating cluster of the single dimerized layer is quantum-critical on the whole line 
$(0 < g \le g^*,p^*)$, with $g^* \approx 1.25$. On this line, the magnetization fluctuations 
are not completely quenched as $T \to 0$, leading to a divergent  susceptibility, 
$\chi \sim T^{-\alpha}$, with $\alpha \to 1/2$ for $g \to g^{*-}$ and $\alpha \to 1$ for
$g \to 0$. For $g=0$, a Curie susceptibility is expected on account of the percolating cluster 
breaking up into smaller pieces when all the $J_2$ couplings vanish (the percolation point here 
is $p^*_0 \approx 0.29$). Some of these clusters contain an odd number of spins. For $g > 0$, 
the divergent susceptibility can then be attributed to effectively isolated subclusters with 
net moments, which are gradually ``frozen out" as the temperature is reduced. The form 
$\chi \sim T^{-\alpha}$ corresponds to a free-moment density scaling as $T^{1-\alpha}$. This 
remarkable behavior will here be demonstrated on the basis of large scale QMC (stochastic 
series expansion \cite{sse}) calculations. Only results exactly at the percolation 
threshold, $p=p^*$, will be discussed. The numerical techniques and special methods developed 
for studies of random systems at ultra-low temperature are discussed in detail in \cite{perc}.

\begin{figure}
\includegraphics[width=5.5cm, clip]{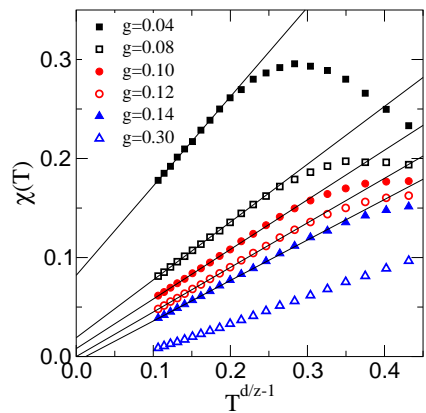}
\caption{(Color online) Temperature scaling of the susceptibility of the infinite 
percolating bilayer cluster for different coupling ratios $g$, using a dynamic exponent 
$z=1.36$.}
\label{fig2}
\end{figure}

The temperature dependence of the uniform susceptibility $\chi(T)$ of the bilayer close
to the multi-critical point was discussed before in Ref.~\cite{sandvik1} (averages over all
clusters were presented in \cite{vajk2}). Fig.~\ref{fig2} shows a more extensive set of
high-precision results for the largest cluster on $L\times L$ lattices with $L=256$ ($L \to \infty$ 
converged for the temperatures shown) at $p^* = 0.4072538$ \cite{newman}. Averages over several 
thousand dilution realizations were taken. The temperature is scaled according to the expected 
quantum-critical form, $\chi = a + bT^{d/z-1}$ \cite{chubukov}, where $a$ and $b$ are constants 
and $a=0$ at a quantum-critical point. Using the fractal dimension $d=91/48$  
and adjusting $z$ to obtain a linear $\chi$ versus $T^{d/z-1}$, the same dynamic 
exponent, $z = 1.36 \pm 0.01$, is found for all $0 < g \alt g^*$. An improved estimate $g^* = 0.118 
\pm 0.006$ is also obtained. Note that bilayer criticality can be expected only for 
$T < g$, which is indeed the case in Fig.~\ref{fig2}.

In a clean quantum-critical system, there is a low-temperature cross-over of $\chi(T)$ to a 
``renormalized classical" behavior when $g < g_c$, at a temperature of the order of the spin 
stiffness $\rho_s$ \cite{chubukov}. No such cross-over is seen in Fig.~\ref{fig2}, however. 
Although it cannot be completely excluded that there is a cross over at still lower temperature, 
one can also argue that there should be no cross-over, because $\rho_s=0$ on the  percolating 
cluster \cite{perc,sandvik1} (although there is long-range order---this unusual behavior
has also been discussed in \cite{brayali1}) and there is no apparent 
energy scale, except $T$, to govern the long-distance physics of the spins on this fractal network. 
Thus the multi-critical point $(p^*,g^*)$ may control the $T < g$ temperature dependence for all 
$0 \le g \le g^*$ at $p^*$. Considering that the scaling in Fig.~\ref{fig2} extends down as low as to 
$T = J/256$ and to couplings $g \approx g^*/3$, the results appear to support such an unusual 
manifestation of quantum-criticality on a fractal percolating cluster.

\begin{figure}
\includegraphics[width=5.9cm, clip]{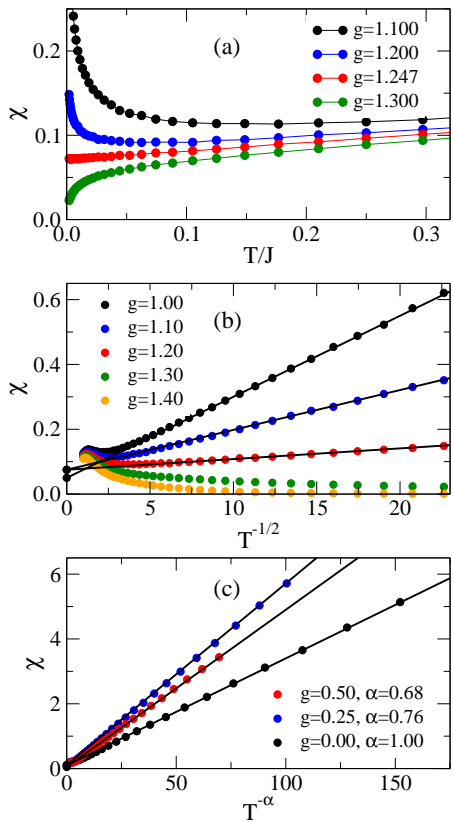}
\caption{(Color online) Temperature dependence of the susceptibility of the percolating 
single-layer cluster. (a) At and close to the coupling $g^* \approx 1.247$. (b) $T^{-1/2}$ scaling
for $1 < g \alt g^*$. (c) Scaling with varying exponent for $g < 1$.}
\label{fig3}
\end{figure}

Turning now to the single-layer model at its dimer percolation point, $p^*=1/2$ for $g>0$, 
the disorder-averaged uniform susceptibility of the largest cluster was calculated for 
$L$ up to $256$, down to $T=J/512$ ($L \to \infty$ converged). Here a special point 
$(g^*,p^*)$ is found which separates qualitatively different
behaviors of the uniform susceptibility. As seen in Fig.~\ref{fig3}(a), at 
$g^* = 1.247 \pm 0.001$, $\chi$ is linear in $T$ and approaches a finite value as 
$T \to 0$. For $g > g^*$ the susceptibility drops to zero and for $g < g^*$ it diverges.
As shown in Figs.~\ref{fig1}(b),(c), the divergence is of the form $\chi \sim T^{-\alpha}$,
with $\alpha$ very close to $1/2$ for $ 1 \alt g < g^*$ and $\alpha \to 1$ for $g \to 0$. 
Such a divergence can be interpreted as a temperature dependent fraction $\sim T^{1-\alpha}$ 
of effectively free magnetic moments. 

As already noted, exactly at $g=0$ the percolating cluster is broken up into smaller subclusters 
and then a Curie behavior, $\chi \sim T^{-1}$, is expected on account of clusters with a net spin. 
For small but non-zero $g$, one might then have expected a cross-over from Curie behavior when 
$T > g$ to a finite susceptibility as $T\to 0$. Instead, it appears that coupling the subclusters 
leads to a $g$ dependent power-law temperature scaling of the number of effectively free moments.
The effective couplings of these moments to each other must thus have a 
$g$ dependent power-law distribution, leading to a self-similar structure of free moments different 
from that of the underlying fractal cluster. It is remarkable that this behavior persists also when 
$g \approx 1$, where the picture of weakly connected subclusters is not obviously relevant.

\begin{figure}
\includegraphics[width=6.5cm, clip]{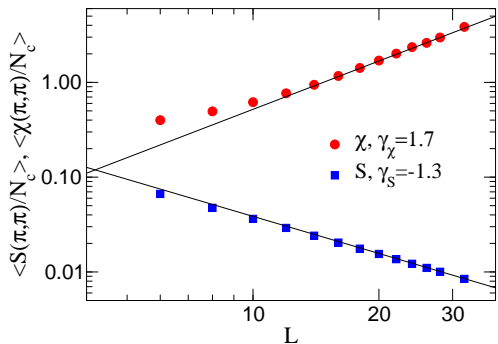}
\caption{(Color online)
Finite-size scaling of the staggered structure factor and susceptibility, 
normalized by the cluster size, at intra-dimer coupling $g=1$. The lines show scaling with 
exponents indicated in the legends. The results where obtained at $T$ sufficiently low 
to give the ground state.}
\label{fig4}
\end{figure}

The bilayer percolating cluster is ordered at $T=0$ for $g < g^*$ \cite{sandvik1,vajk2}. 
This is not the case for the single dimerized layer at $p^*$. Instead, quantum-critical 
fluctuations are observed for $0 < g \le g^*$. Consider the staggered structure factor 
$S(\pi,\pi)$ and susceptibility $\chi(\pi,\pi)$ of a cluster of $N_c$ spins,
\begin{eqnarray}
S(\pi,\pi) & = & \frac{1}{N_c} \sum_{i,j}P_{ij} \langle S^z_i S^z_j \rangle, \nonumber \\
\chi(\pi,\pi) & = & \frac{1}{N_c} \sum_{i,j} P_{ij} \int_0^\beta d\tau 
\langle S^z_i(\tau) S^z_j(0) \rangle, \nonumber
\end{eqnarray}
where $P_{ij}=(-1)^{x_i+y_i-x_j-y_j}$. At a quantum-critical point, these quantities, 
averaged over disorder, should scale with the system size as 
$\langle S(\pi,\pi)/N_c \rangle \sim L^{\gamma_s}$, 
$\langle \chi(\pi,\pi)/N_c \rangle \sim L^{\gamma_\chi}$, with $\gamma_s = -(z+\eta)$ and 
$\gamma_\chi = -\eta$ (normalizing by $N_c \sim L^d$ before disorder-averaging leads to some 
reduction of statistical fluctuations). Fig.~\ref{fig4} shows results for $g=1$. The observed 
scaling gives $z \approx 3.0$ and $\eta \approx -1.7$. The dynamic exponent can be compared with 
the expected quantum-critical susceptibility; $\chi \sim T^{-\alpha}$ with $\alpha = 1- d/z$ 
\cite{chubukov}. With the exponent $\alpha=1/2$ obtained from $\chi (T)$ in Fig.~\ref{fig3}(b), 
it is apparent that this relationship does not hold here [$z$ extracted from $S(\pi,\pi)$ and
$\chi(\pi,\pi)$ should be the actual dynamic exponent]. At the special point $g^*$ the 
extracted $\alpha=0$ and $z\approx d$ (not shown here) are in fact consistent with this 
relationship. This is also the case at the bilayer multi-critical point \cite{sandvik1}. 
The single-layer quantum-criticality on the line $( 0 < g < g^*,p^*)$ thus appears to be 
fundamentally different. It should be noted that the behavior is not consistent with
a Griffiths phase \cite{lin}, as the spin correlations in that case should be exponentially 
decaying, in contrast to the power-law seen in Fig.~\ref{fig4}.

An extended line of quantum-critical points was not anticipated on the basis 
of a real space RG approach developed recently for quantum rotors on a percolating cluster 
\cite{brayali}. The critical line discovered here for the single-layer model is more
similar to the 1D random singlet phase \cite{fisher}, in that there is a temperature 
dependent fraction of effectively free moments. In the random singlet phase 
$z=\infty$ whereas in the model studied here $z$ is finite and diverges in the 
limit $g \to 0$. A temperature dependent fraction of effective moments has also been 
observed in a 2D model of interacting localized moments \cite{laflorencie}. However, there 
the asymptotic $T\to 0$ susceptibility is always Curie-like, and there is long-range 
order at $T=0$. An RG calculation for random frustrated moments in the continuum shows 
a $\chi(T)$ divergence with varying exponent \cite{bhatt}. The ground state properties 
were not accessible in that study.

The bilayer multi-critical point $(p^*,g^*)$ has been argued \cite{sandvik1,vajk2} to influence 
finite-temperature properties of single-layer Zn doped cuprate antiferromagnets, for which a dynamic 
exponent $z\approx 1.4$ was found in neutron scattering experiments \cite{vajk}. However, although 
$g^*$ is small ($\approx 0.12$) it is difficult to explain how bilayer quantum-criticality could 
be realized when $T \gg g =0$ (due to an expected cross-over at $T \approx g$; see Fig.~\ref{fig2}). 
Physical realizations of single-layer dimer dilution are not immediately obvious. Nevertheless, 
the results presented here serve to illustrate rich and surprising behaviors arising from the 
interplay of classical percolation and quantum fluctuations, going beyond previous examples of 
scaling in percolating fractal structures \cite{orbach}.

\acknowledgments{This work was supported by the NSF under Grant No.~DMR05-13930.}

\null


\vskip-10mm
\begin{thebibliography}{00}

\bibitem{sachdev} 
S. Sachdev, {\it Quantum Phase Transitions} (Cambridge University Press,
Cambridge 1999).

\bibitem{chn} 
S. Chakravarty, B. I. Halperin, and D. R. Nelson, Phys. Rev. Lett. {\bf 60},
1057 (1988); Phys. Rev. B {\bf 39}, 2344 (1989).

\bibitem{chubukov} 
A. V. Chubukov, S. Sachdev and J. Ye, Phys. Rev. B {\bf 49}, 11919 (1994).

\bibitem{ma}
S.-K. Ma, C. Dasgupta, and C.-K. Hu, Phys. Rev. Lett. {\bf 43}, 1434 (1979);
C. Dasgupta. and S.-K. Ma, Phys. Rev. B {\bf 22}, 1305 (1980).

\bibitem{fisher}
D. S. Fisher, Phys. Rev. B {\bf 50}, 3799 (1994).

\bibitem{pich}
C. Pich, A. P. Young, H. Rieger, and N. Kawashima, Phys. Rev. Lett. {\bf 81}, 5916 (1998).

\bibitem{lin}
Y.-C. Lin, R. M\'elin, H. Rieger, and F. Igl\'oi, Phys. Rev. B {\bf 68},
024424 (2003); Y.-C. Lin, H. Rieger, N. Laflorencie, and F. Igl\'oi, cond-mat/0604126.

\bibitem{lafl}
N. Laflorencie, S. Wessel, A. Laeuchli, and H. Rieger, Phys. Rev. B {\bf 73}, 060403(R) (2006).

\bibitem{troyer}
M. Troyer, M. Imada, and K. Ueda, J. Phys. Soc. Jpn. {\bf 66}, 2957 (1997).   

\bibitem{matsumoto}
M. Matsumoto, C. Yasuda, S. Todo, and H. Takayama, Phys. Rev. B {\bf 65}, 014407 (2001).

\bibitem{bilayqmc}
L. Wang, K. S. D. Beach, and A. W. Sandvik, Phys. Rev. B {\bf 73}, 014431  (2006).

\bibitem{haldane} 
F. D. M. Haldane, Phys. Rev. Lett. {\bf 61}, 1029 (1988).

\bibitem{harris}
A. B. Harris, J. Phys. C {\bf 7}, 1671 (1974).

\bibitem{sse}
A. W. Sandvik, Phys. Rev. B {\bf 59}, R14157 (1999).

\bibitem{sandvik1}
A. W. Sandvik, Phys. Rev. Lett. {\bf 89}, 177201 (2002).

\bibitem{vajk2}
O. P. Vajk and M. Greven, Phys. Rev. Lett. {\bf 89}, 177202 (2002).

\bibitem{sknepnek}
R. Sknepnek, T. Vojta, and M. Vojta, Phys. Rev. Lett. {\bf 93}, 097201 (2004).

\bibitem{yu}
R. Yu, T. Roscilde, and S. Haas, Phys. Rev. lett. {\bf 94}, 197204 (2005).

\bibitem{monthus}
F. Igl\'oi and C. Monthus, Phys. Rep. {\bf 412}, 277  (2005).

\bibitem{carretta}
P. Carretta, A. Rigamonti, and R. Sala, Phys. Rev. B {\bf 55}, 3734 (1997).

\bibitem{vajk}
O. P. Vajk, P. K. Mang, M. Greven, P. M. Gehring, and J. W. Lynn,
Science {\bf 295}, 1691 (2002).

\bibitem{wan}
C. C. Wan, A. B. Harris, and J. Adler, J. Appl. Phys. {\bf 69}, 5191 (1991).

\bibitem{yasuda}
C. Yasuda and A. Oguchi, J. Phys. Soc. Jpn. {\bf 66}, 2836 (1997);
J. Phys. Soc. Jpn., {\bf 68} 2773 (1999).

\bibitem{chen}
Y.-C. Chen and A. H. Castro Neto, Phys. Rev. B {\bf 61}, R3772 (2000).

\bibitem{kato}
K. Kato, S. Todo, K. Harada, N. Kawashima, S. Miyashita, and H. Takayama,
Phys. Rev. Lett. {\bf 84}, 4204 (2000).

\bibitem{perc}
A. W. Sandvik, Phys. Rev. B {\bf 66}, 024418 (2002).

\bibitem{vojta}
T. Vojta and J. Schmalian, Phys. Rev. Lett. {\bf 95}, 237206 (2005).

\bibitem{stauffer}
D. Stauffer and A. Aharony, {\it Introduction to Percolation Theory}
(Taylor and Francis, London, 1991).

\bibitem{singledil}
N. Nagaosa, A. Furusaki, M. Sigrist, and H. Fukuyama, J. Phys. Soc. Jpn. 
{\bf 65}, 3724 (1996).

\bibitem{senthil}
T. Senthil and S. Sachdev, Phys. Rev. Lett. {\bf 77}, 5292 (1996).

\bibitem{brayali}
N. Bray-Ali, J. E. Moore, T. Senthil, and A. Vishwanath, 
Phys. Rev. B {\bf 73}, 064417 (2006).

\bibitem{newman}
M. E. J. Newman and R. M. Ziff, Phys. Rev. Lett. {\bf 85}, 4104 (2000).

\bibitem{brayali1}
N. Bray-Ali, J. E. Moore, Phys. Rev. B {\bf 69}, 184505 (2004).

\bibitem{laflorencie}
N. Laflorencie, D. Poilblanc, and M. Sigrist, Phys. Rev. B {\bf 71}, 212403 (2005).

\bibitem{bhatt}
R.\ N.\ Bhatt and P.\ A.\ Lee, Phys. Rev. Lett. {\bf 48}, 344 (1982).

\bibitem{orbach}
T.\ Nakayama, K.\ Yakubo, and R.\ Orbach, Rev. Mod. Phys. {\bf 66}, 381 (1994).

\end{thebibliography}
\end{document}